\documentclass[aps,graphicx,twocolumn]{revtex4}

\usepackage{graphicx}
\usepackage{epstopdf}
\usepackage{verbatim}
\usepackage{amsmath}
\usepackage[colorlinks,linkcolor=blue,citecolor=blue,urlcolor=blue]{hyperref}
\usepackage[caption=false,font=scriptsize,labelfont=sf,textfont=sf]{subfig}

\begin{document}
\title{Device-independent quantum secret sharing with noise preprocessing and postselection}

\author{Qi Zhang$^{1}$, Wei Zhong$^{3}$, Ming-Ming Du$^{2}$, Shu-Ting Shen$^{2}$, Xi-Yun Li$^{1}$, An-Lei Zhang$^{1}$, Lan Zhou$^{1}$\footnote{Email address: zhoul@njupt.edu.cn}, Yu-Bo Sheng$^{2}$\footnote{Email address: shengyb@njupt.edu.cn}}
\address{
$^1$College of Science, Nanjing University of Posts and Telecommunications, Nanjing, Jiangsu 210023, China\\
$^2$College of Electronic and Optical Engineering and College of Flexible Electronics (Future Technology), Nanjing University of Posts and Telecommunications, Nanjing, Jiangsu 210023, China\\
$^3$Institute of Quantum Information and Technology, Nanjing University of Posts and Telecommunications, Nanjing, Jiangsu 210003, China\\
}
\date{\today}

\begin{abstract}
Device-independent (DI) quantum secret sharing (QSS) can relax the security assumptions about the devices' internal workings and provide QSS the highest level of security in theory. The original DI QSS protocol proved its correctness and completeness under a causal independence assumption regarding measurement devices. However, there has been a lack of DI QSS's performance characterization in practical communication situations, which impedes its experimental demonstration and application in the future. Here, we propose a three-partite DI QSS protocol with noise preprocessing and postselection strategies and develop the numerical methods to implement its performance characterization in practical communication situations. The adoption of the noise preprocessing and postselection can reduce DI QSS's global detection efficiency threshold from 96.32\% to 94.30\% and increase the noise threshold  from 7.148\% to 8.072\%. Our DI QSS protocol has two advantages. First, it is a DI QSS protocol with performance characterization in practical communication situations. Second, the adoption of noise preprocessing and postselection can effectively relax its experimental requirement and enhance the noise resistance. Our DI QSS protocol has potential for future experimental demonstration and application.
\end{abstract}
\maketitle

\section{Introduction}
Quantum secure communication is based on the basic principles of quantum mechanics to guarantee the unconditional security of the transmitted messages. Quantum secure communication mainly includes three important branches, specifically, quantum key distribution (QKD) \cite{QKD1,QKD2,QKD3n,QKD2n,QKD4n,QKD4}, quantum secure direct communication \cite{QSDC2,QSDC3,QSDC4}, and quantum secret sharing (QSS) \cite{QSS1,QSS6,QSS7}. Quantum key distribution is used to distribute secret keys between two distant parties. Quantum secret sharing is a multipartite cryptographic primitive. It aims to split a secret key of one user, called the dealer, into several parts and distribute each part to a user, called the player. Any unauthorized subset of players cannot reconstruct the distributed key, which can be reconstructed only when all the authorized players cooperate \cite{QSS1}. Quantum secret sharing has important applications in many quantum information tasks, such as quantum Byzantine agreements \cite{agreement} and distributed quantum computation \cite{computing}.

The first QSS protocol was proposed in \cite{QSS1}. Since then, QSS has been widely researched in both theory and experiment \cite{QSS7n,QSS6,QSS7,QSS8,QSS8n,QSS11,QSS12,QSS13,QSS10,QSS10n,MDIQSS,MDIQSS1,DMDIQSS1,DPS2,RR1,QSS14,QSSe2,QSSe4,QSSe6,QSSe6n,QSSe3,QSSe5,QSSe5n,QSSe7,QSSe8,QSSe9}. The Greenberger-Horne-Zeilinger (GHZ) state, with its multibody entanglement property, is a common quantum resource of QSS. Chen \emph{et al.} \cite{QSSe2} developed and utilized an ultrastable high-intensity four-photon GHZ state source to experimentally demonstrate a three-user QSS. During the past few decades, experiments of QSS based on entanglement \cite{QSSe2,QSSe4,QSSe6,QSSe6n}, a single qubit \cite{QSSe3}, a graph state \cite{QSSe5,QSSe5n}, and a coherent state \cite{QSSe7} have been reported. Similar to other quantum secure communication branches, practical imperfect experimental devices may cause security loopholes for QSS. In \cite{MDIQSS} a measurement-device-independent QSS protocol was proposed, which can resist all possible attacks from imperfect measurement devices. However, the side message leakage caused by the practical imperfect photon source may still threaten the QSS's security \cite{BB841,BB842,BB843}. A promising approach to eliminate the security loophole from all practical imperfect devices is to use a device-independent (DI) method. Device-independent-type protocols only require two fundamental assumptions, that the quantum physics is correct and no unwanted signal can escape from each party's laboratory. They treat all experimental devices in each user's location as a black box and eliminate all additional assumptions for the experimental devices. Device-independent-type protocols can resist all possible attacks from practical experimental devices and provide the highest security \cite{DIQKD,DIQKD1,DIQKD2,DIQRG1,DIQRG2}.

The study on DI-type protocols started with DI QKD \cite{DIQKD,DIQKD1,DIQKD2}. During past few years, DI QKD has achieved great development in theory \cite{DIQKD3,DIQKD4,DIQKD5,DIQKD14,DIQKD5n,DIQKD8,DIQKD9,DIQKD10,DIQKD11,DIQKD12,DIQKD7,DIQKD13,DIQKD15,DIQKD16,DIQKD17}. The security of DI QKD is based on two-particle nonlocal correlations, which can be detected by the violation of Bell-type inequalities [typically, the Clauser-Horne-Shimony-Holt (CHSH) inequality] \cite{Bell,CHSH}. Device-independent QKD made great breakthroughs in experiments in \cite{DIQKDe1,DIQKDe2,DIQKDe3}. Device-independent-type protocols require a large number of signal pairs that are sufficiently strongly entangled. The photon loss and decoherence caused by the imperfect devices and channel noise would seriously destroy the entanglement. In this way, DI-type protocols require quite high global detection efficiency (over 90\%) and have a low noise threshold \cite{DIQKD1,DIQKD2,DIQKD3}, which makes them elusive to realize with current technologies, especially in optical implementations. In the DI QKD field, researchers have used some active improvement strategies to relax DI QKD's global detection efficiency requirement and enhance its noise threshold, such as noise preprocessing and postselection \cite{DIQKD5n,DIQKD14,DIQKD8,DIQKD9,DIQKD10,DIQKD11,DIQKD12,DIQKDe3}. Recently, the adoption of a noise preprocessing strategy increased DI QKD's noise threshold from an initial 7.1492\% to 8.0848\% and reduced the global detection efficiency threshold from 92.4\% to 90.78\% \cite{DIQKD9,DIQKD11}. The DI QKD's global detection efficiency threshold was further reduced to less than 87.49\% by combining random postselection and noise preprocessing strategies, and the first DI QKD experiment in an optical platform was demonstrated \cite{DIQKDe3}.

Roy and Mukhopadhyay proposed the first DI QSS protocol in arbitrary even dimensions and proved its correctness and completeness under a causal independence assumption
regarding measurement devices \cite{DIQSS2}. A stronger form of Bell nonlocality was proposed in \cite{DIQSS1} that could avoid possible attacks in secret sharing. However, so far, there has been a lack of DI QSS's performance characterization in practical communication situations, which impedes its experimental demonstration and applications in the future. In the paper, we propose a three-partite DI QSS protocol with noise preprocessing and postselection. We develop the numerical methods to implement its performance characterization in practical communication situations, including the key generation rate via von Neumann entropy, the global detection efficiency, the noise threshold, and the maximal communication distance between any two users. Moreover, the adoption of noise preprocessing and postselection can effectively relax DI QSS's global detection efficiency requirement and enhance its noise resistance. Based on the above features, our DI QSS protocol has the potential for experimental demonstration and applications in the future quantum secure communication field.

The paper is organized as follows. In Sec. \ref{Section2} we explain the genuine tripartite nonlocal correlation, which is the core of our DI QSS protocol. In Sec. \ref{Section3} we propose the DI QSS protocol without any active improvement and estimate its performance in a practical channel environment. In Sec. \ref{Section4} we adopt noise preprocessing and postselection strategies in the DI QSS protocol and estimate its performance. In Sec. \ref{Section5} we summarize our work and discuss the results.

\section{Tripartite nonlocal correlation}\label{Section2}
Before explaining our DI QSS protocol, we briefly introduc genuine tripartite nonlocality, which represents the strongest form of tripartite nonlocality. Svetlichny proposed the first method to detect genuine tripartite nonlocality \cite{nonlocality3}. Svetlichny derived a Bell-type inequality for a three-qubit system, the Svetlichny inequality, which holds even if (any) two of the three parts can display arbitrary nonlocal correlations while the third party is separated \cite{nonlocality3,nonlocality5}.

Suppose that three separated parties Alice, Bob, and Charlie perform measurement on their photons. We denote the measurement basis of Alice, Bob, and Charlie by $A_i$, $B_j$, and $C_k$, respectively, where $i,k\in \{1,2\}$. Here, to apply the Svetlichny inequality in our DI QSS protocol, we choose $j\in \{1,2,3\}$. Their measurement results are $a_i$, $b_j$, $c_k\in \{-1,+1\}$. The Svetlichny inequality is written as
\begin{eqnarray}\label{Svetlichnypolynomial}
S_{ABC}&=&\langle a_{1}b_{2}c_{2}\rangle+\langle a_{1}b_{3}c_{1}\rangle+\langle a_{2}b_{2}c_{1}\rangle-\langle a_{2}b_{3}c_{2}\rangle \nonumber\\
&+&\langle a_{2}b_{3}c_{1}\rangle+\langle a_{2}b_{2}c_{2}\rangle+\langle a_{1}b_{3}c_{2}\rangle-\langle a_{1}b_{2}c_{1}\rangle\leq4.\nonumber\\
\end{eqnarray}
Here, $\langle a_{i}b_{j}c_{k}\rangle$ represents the expected value of the tripartite measurement results, $\langle a_{i}b_{j}c_{k}\rangle=P(a_{i}b_{j}c_{k}=1)-P(a_{i}b_{j}c_{k}=-1)$ ($P$ represents the probability). The violation of Svetlichny inequality implies the presence of genuine tripartite nonlocality.

Based on Ref. \cite{nonlocality7}, the Svetlichny polynomial $S_{ABC}$  can be simplified by the CHSH polynomials as
\begin{eqnarray}\label{Svetlichnypolynomial2}
S_{ABC}=\langle S_{AB}c_{2}\rangle+\langle S_{AB}'c_{1}\rangle,
\end{eqnarray}
where $\langle S_{AB}c_{2}\rangle=P(S_{AB}c_{2}=1)-P(S_{AB}c_{2}=-1)$ and $\langle S_{AB}'c_{1}\rangle=P(S_{AB}'c_{1}=1)-P(S_{AB}'c_{1}=-1)$. $S_{AB}$, with $S_{AB}$ the usual CHSH polynomial between Alice and Bob in the form
\begin{eqnarray}\label{CHSH}
S_{AB}=\langle a_{1}b_{2}\rangle+\langle a_{2}b_{2}\rangle+\langle a_{1}b_{3}\rangle-\langle a_{2}b_{3}\rangle,
\end{eqnarray}
where $\langle a_{i}b_{j}\rangle=P(a_{i}b_{j}=1)-P(a_{i}b_{j}=-1)$. By applying the mappings $b_{2}\mapsto b_{3}$ and $b_{3}\mapsto -b_{2}$, which do not affect the local correlation, we can transform $S_{AB}$ to its equivalent form  $S_{AB}'$ as
\begin{eqnarray}\label{CHSH2}
S_{AB}'=\langle a_{2}b_{3}\rangle+\langle a_{2}b_{2}\rangle+\langle a_{1}b_{3}\rangle-\langle a_{1}b_{2}\rangle.
\end{eqnarray}

Assume that Alice, Bob, and Charlie share a three-photon polarization Greenberger-Horne-Zeilinger state in the form
\begin{eqnarray}\label{AGHZ}
|GHZ\rangle=\frac{1}{\sqrt{2}}(|HHH\rangle+|VVV\rangle),
\end{eqnarray}
where $|H\rangle$ and $|V\rangle$ denote the horizontal and vertical polarization states, respectively. To prepare a state for Alice and Bob that is optimal for the corresponding CHSH test, Charlie has two measurement bases $C_{1}=\sigma_x$ and $C_{2}=-\sigma_y$ to measure his photons, where $\sigma_x=\{|+_x\rangle=\frac{1}{\sqrt{2}}(|H\rangle+|V\rangle), |-_x\rangle=\frac{1}{\sqrt{2}}(|H\rangle-|V\rangle) \}$, and $-\sigma_y=\{|+_y\rangle=\frac{1}{\sqrt{2}}(|H\rangle-i|V\rangle), |-_y\rangle=\frac{1}{\sqrt{2}}(|H\rangle+i|V\rangle)\}$. Charlie defines that $|+_x\rangle$ and $|+_y\rangle$ correspond to the measurement result $+1$ and that $|-_x\rangle$ and $|-_y\rangle$ correspond to the measurement result $-1$.

Charlie announces his measurement basis and measurement result for each photon. Under the case that Charlie chooses $C_1=\sigma_{x}$ to measure his photon, when he obtains the measurement result of $\pm1$, the quantum state shared by Alice and Bob will collapse to
\begin{eqnarray}\label{ABell1}
|\phi^{\pm}\rangle_{AB}=\frac{1}{\sqrt{2}}(|HH\rangle\pm|VV\rangle),
\end{eqnarray}
respectively. This is optimal for the CHSH test between Alice and Bob. Alice and
Bob choose measurements which are optimal for the CHSH test, i.e., $A_{1}=\sigma_x$, $A_{2}=\sigma_{y}$, $B_{1}=\sigma_x$, $B_{2}=(\sigma_x-\sigma_{y})/\sqrt{2}$, and $B_{3}=(\sigma_x+\sigma_{y})/\sqrt{2}$. The states $|\phi^{\pm}\rangle_{AB}$ provide the maximal value of the CHSH polynomial of $S_{AB}=\pm2\sqrt{2}$, respectively.

In the case that Charlie chooses $C_2=-\sigma_y$ to measure his photon, when he obtains the measurement result of $\pm1$, the quantum state shared by Alice and Bob will collapse to
\begin{eqnarray}\label{ABell1}
|\psi^{\pm}\rangle_{AB}=\frac{1}{\sqrt{2}}(|HH\rangle\pm i|VV\rangle).
\end{eqnarray}
Similar to the above case, the state $|\psi^{\pm}\rangle_{AB}$ can also lead to the maximal value of $S_{AB}=\pm2\sqrt{2}$, respectively.

In this way, when $c_k=+1$, the measurements of Alice and Bob are positively correlated, with $S_{AB}=2\sqrt{2}$. Conversely, when $c_k=-1$, the measurements of Alice and Bob are anti-correlated, with $S_{AB}=-2\sqrt{2}$. Thus, the maximal violation of the Svetlichny inequality $S_{ABC}=4\sqrt{2}$ can be achieved. Charlie's measurements consistently keep Eq.~\eqref{Svetlichnypolynomial2} positive. Without loss of correlation between Alice and Bob, we use $S$ to represent $S_{AB}$ in the following parts, which is given by
 \begin{equation}\label{Bell2}
	S=\begin{cases}
	   S_{AB}'& \text{if     $c_1=+1$},\\
	   -S_{AB}'& \text{if     $c_1=-1$},\\
        S_{AB}& \text{if     $c_2=+1$},\\
	   -S_{AB}& \text{if     $c_2=-1$}.
	\end{cases}
\end{equation}
The above simplification shows that the Svetlichny polynomial can be treated as an extension of the CHSH polynomial in the tripartite scenario. In this way, the genuine tripartite nonlocality of the three photons can be guaranteed by Alice's and Bob's results violating the CHSH inequality ($S\leq2$).

\section{DI QSS protocol}\label{Section3}
In this section, we describe the DI QSS protocol in detail and estimate its performance in practical quantum communication situations. Similarly to DI QKD protocols \cite{DIQKD,DIQKD1,DIQKD2}, DI QSS's security can also be guaranteed by two fundamental assumptions, specifically that the quantum physics is correct and the users' physical locations are secure. Meanwhile, the three users must be legitimate and honest during the key generation process. The basic principle of our DI QSS protocol is shown in Fig. \ref{fig:boat1}. The DI QSS protocol includes six steps, which are shown below.

\subsection{DI QSS process}\label{Section3.1}
\begin{figure}[t]
\includegraphics[scale=0.43]{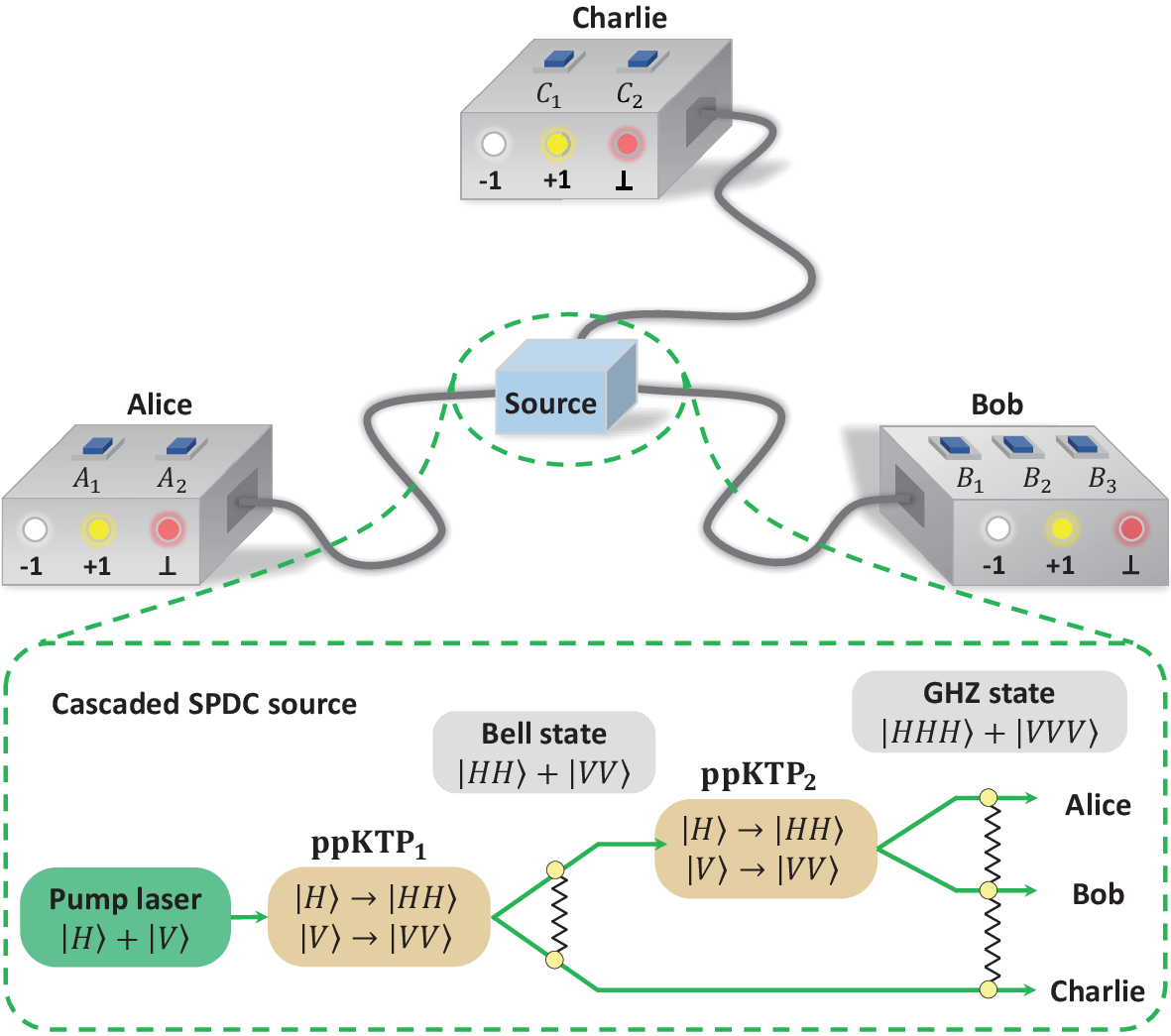}
\caption{(Color online) Schematic diagram of the DI QSS protocol. The center source based on cascaded SPDC prepares a large number of identical three-photon polarization GHZ states, which are  divided into three sequences, namely, $E_1$, $E_2$, and $E_3$ sequences. The photons in three sequences are distributed to three distant users, Alice, Bob, and Charlie, respectively. The three users select a measurement basis independently and get the measurement results $-1$, $+1$ or no-click event $\perp$.}
\label{fig:boat1}
\end{figure}

\emph{Step 1} The central entanglement source generates $N$  pairs of three-photon polarization GHZ states in the form of Eq.~\eqref{AGHZ} based on the cascaded spontaneous covariant down-conversion (SPDC) process \cite{GHZ2} ($N$ is a large number). The central entanglement source divides the three photons of each GHZ state into $E_1$, $E_2$, and $E_3$ sequences, respectively. Then all the photons in $E_1$, $E_2$, and $E_3$ sequences are distributed to three distance users, Alice, Bob, and Charlie sequentially through three quantum channels, respectively. Here, we consider a symmetrical situation, i.e., the distance from the entanglement source to each user is equal.

\emph{Step 2} After three users receive the photons, they independently and randomly select a measurement basis to measure the received photons. Both Alice and Charlie have two measurement bases, i.e., $A_1=\sigma_x$ and $A_2=\sigma_y$ for Alice and  $C_1=\sigma_x$ and $C_2=-\sigma_y$ for Charlie. Bob has three measurement bases $B_1=\sigma_x$, $B_2=(\sigma_x-\sigma_y)/\sqrt{2}$, and $B_3=(\sigma_x+\sigma_y)/\sqrt{2}$. The measurement results from the three users are denoted by $a_{i}$, $b_{j}$, and $c_{k}$, respectively ($i,k\in \{1,2\}$ and $j\in \{1,2,3\}$), where $a_{i}$, $b_{j}$, $c_{k}\in \{-1,+1\}$. We also assume that each user's measurement result is only a function of the current inputs.

\emph{Step 3} After the measurement, Alice, Bob, and Charlie announce the measurement basis for each photon in $E_1$, $E_2$, and $E_3$ sequences in turn. There are three cases based on their measurement basis selection.

In the first case, Bob chooses $B_2$ or $B_3$ (corresponding to eight measurement basis combinations $\{A_1 B_2 C_1\}$, $\{A_1 B_2 C_2\}$, $\{A_1 B_3 C_1\}$, $\{A_1 B_3 C_2\}$, $\{A_2 B_2 C_1\}$, $\{A_2 B_2 C_2\}$, $\{A_2 B_3 C_1\}$, and $\{A_2 B_3 C_2\}$). All the users announce their measurement results, which are used to estimate the Svetlichny polynomial $S_{ABC}$ (the CHSH polynomial $S$) for the security checking (the detailed estimation process is shown in Sec. \ref{Section2}). When $S_{ABC}$ reaches the maximal value of $4\sqrt{2}$ (which is equivalent to $S$ reaching the maximal value of $2\sqrt{2}$), the users share the maximally entangled GHZ state. This ideal case corresponds to the perfect quantum channel and experimental devices. Any eavesdropping behavior from Eve will break the nonlocality of the corresponding measurement results and thus reduce the value of $S_{ABC}$ ($S$) in statistics. As long as the users detect the reduction of $S_{ABC}$ ($S$), they can  detect the eavesdropping, and thus the key leakage rate  is strictly zero. In practical experimental conditions, the imperfect quantum channel and experimental devices may reduce $S_{ABC}$ ($S$). When $S_{ABC}>4$ (which is equivalent to $S>2$), the corresponding measurement results of the three users are still nonlocally correlated. In this scenario, Eve can steal some transmitted keys, but the users can bound the key leakage rate to Eve. In this way, the users still regard the photon transmission process to be secure and the communication continues. In contrast, when $S_{ABC}\leq4$  (which is equivalent to $S\leq2$), the three users' measurement results are classically correlated. In this scenario, the users cannot detect Eve's eavesdropping, so the photon transmission process is not secure. The users have to terminate the communication.

In the second case, when the measurement basis combination is $\{A_{1}B_{1}C_{1}\}$, the users retain the measurement results as the raw key bits. Each party's measurement result $+1$ is labeled as the key bit 0, and $-1$ is labeled as the key bit 1. The encoding rule can be described as $k_{A}=k_{B}\oplus k_{C}$, where $k_A$, $k_B$ and $k_C$ denote the key bits of Alice, Bob, and Charlie, respectively. Alice randomly announces some of the raw key bits, while Bob and Charlie announce the corresponding raw key bits to estimate the quantum bit error rate (QBER) $\delta$. They preserve the remaining unpublished raw key bits.

In the third case, the measurement basis combination is $\{A_{1}B_{1}C_{2}\}$, $\{A_{2}B_{1}C_{1}\}$, or $\{A_{2}B_{1}C_{2}\}$. In this case, the users have to discard their measurement results.

\emph{Step 4} The users repeat the above steps until they obtain a sufficient number of raw key bits.

\emph{Step 5} The users perform the error correction and private amplification on the obtained raw keys, resulting in a series of secure key bits.

{\emph{Step 6} Charlie announces his key bit $k_C$ and Bob combines his key bit $k_B$ to reconstruct the key bit $k_A$ delivered by Alice.

\subsection{The performance of the DI QSS protocol in practical communication scenario}
The DI framework aims to guarantee the security of a protocol without specifying the states and measurements used in the protocol. In the DI QSS protocol, we only require that Eve obeys the laws of quantum physics, but do not limit her ability. Eve can even control the entanglement source and fabricate users' measurement devices. Moreover, we assume that Eve has a perfect quantum channel. Although we have specified a particular state (GHZ state) in our DI QSS protocol that produces these nonlocal correlations, we do not assume anything about the source state in its practical implementation. The users only rely on the tripartite nonlocality to certify that the outputs from uncharacterized devices are genuinely random to Eve and thus guarantee the security of the transmitted keys. In the security checking process, the users can only use the relation between the measurement basis selection (input) and measurement results (outcome) to bound Eve's knowledge.

We consider a general attack, i.e., a collective attack, where Eve applies the same attack on each system of Alice and Bob. After the photon transmission, we assume that all the three-photon pairs have the same form. In our DI QSS protocol, we simplify the Svetlichny polynomial $S_{ABC}$ with the combination of the CHSH polynomial $S$. The violation of the Svetlichny inequality for the three users' results is equal to the violation of the CHSH inequality for Alice's and Bob's results. In this way, the security proof of our DI QSS protocol  is similar as that of the DI QKD \cite{DIQKD1,DIQKD2}.

In the DI QSS protocol, when the users' measurement basis combination is $\{A_{1}B_{1}C_{1}\}$, Bob can read out the transmitted key from Alice by combining Charlie's key bit with his own key bit.
We adopt the Devetak-Winter bound \cite{DWrate,DWrate2} to estimate the key generation rate of the DI QSS protocol. The Devetak-Winter bound is a universal method for calculating the key rate in the quantum cryptography field, which has been widely used in QKD and QSS systems \cite{DWrate2,QKD7,QKD8,QKD9,QKD10} and extended to DI QKD and one-side DI QSS systems under the collective attack assumption \cite{QSS10n,DIQKD8,DIQKD16,DIQKD17,DIQKD18}. In the asymptotic limit of a large number of rounds, we conjecture that the key generation rate $r$ of our DI QSS protocol after optimal one-way error correction and privacy amplification is given by \cite{DWrate,DWrate2}
\begin{eqnarray}\label{keyrate}
r &=& I(A_1;B_1,C_1 )-I(A_1;E)\nonumber\\
&=&[H(A_1)-H(A_1|B_1,C_1)]-[H(A_1)-H(A_1|E)]\nonumber\\
&=&H(A_1|E)-H(A_1|B_1,C_1),
\end{eqnarray}
where $I(A_1;B_1,C_1)$ is the mutual information of players Bob and Charlie to the dealer Alice; $I(A_1;E)$ is the mutual information between Alice and Eve; $H(|)$ is the von Neumann conditional entropy; $H(A_1|E)$ quantifies Eve's uncertainty about Alice's key, which can also represent the key secrecy rate to Eve; and $H(A_1|B_1,C_1)$ quantifies the key error rate to Bob and Charlie under the measurement basis of $\{A_{1}B_{1}C_{1}\}$. In practical experiments, $H(A_1|E)$ can be lower bounded by the CHSH polynomial $S$.

During the  long-distance photon transmission process, the photon loss and decoherence caused by the channel noise would seriously destroy the entanglement and weaken the nonlocal correlation among users' measurement results. Here we consider the white-noise model, which is a simple random noise channel model that is widely used in DI QKD protocols for noise theoretical analysis \cite{DIQKD2,DIQKD8,DIQKD9,DIQKD11,DIQKD12}. In the white-noise model, the target GHZ state may degrade to eight possible GHZ states with equal probability. Meanwhile, we assume that each user successfully detects the transmitted photon with a global detection efficiency of $\eta$, and thus the probability of no click is $\bar{\eta}=1-\eta$. In this way, after the entanglement distribution, the users share $N$ pairs of mixed states in the form
\begin{eqnarray}\label{rhoABCnoise}
\rho_{ABC}&=&\eta^3(F|GHZ\rangle\langle GHZ|+\frac{1-F}{8} I)\nonumber\\
&+&\frac{1}{2}\eta^2\bar{\eta}\left(|HH\rangle\langle HH|+|VV\rangle\langle VV|\right)_{BC}\nonumber\\
&+&\frac{1}{2}\eta^2\bar{\eta}\left(|HH\rangle\langle HH|+|VV\rangle\langle VV|\right)_{AC}\nonumber\\
&+&\frac{1}{2}\eta^2\bar{\eta}\left(|HH\rangle\langle HH|+|VV\rangle\langle VV|\right)_{AB}\nonumber\\
&+&\frac{1}{2}\eta\bar{\eta}^2\left(|H\rangle\langle H|+|V\rangle\langle V|\right)_{A}\nonumber\\
&+&\frac{1}{2}\eta\bar{\eta}^2\left(|H\rangle\langle H|+|V\rangle\langle V|\right)_{B}\nonumber\\
&+&\frac{1}{2}\eta\bar{\eta}^2\left(|H\rangle\langle H|+|V\rangle\langle V|\right)_{C}+\bar{\eta}^3|vac\rangle\langle vac|,\nonumber\\
\end{eqnarray}
where the fidelity $F$ is the probability that the photon state is free of error, the unit matrix $I$ consists of the density matrix of the eight possible GHZ states induced by noise (see the Appendix \ref{Appendix A}), and $|vac\rangle$ denotes the vacuum state.

According to the coding rule in step 3, four GHZ states $|GHZ_1^-\rangle=\frac{1}{\sqrt{2}}(|HHH\rangle-|VVV\rangle)$, $|GHZ_2^-\rangle=\frac{1}{\sqrt{2}}(|HHV\rangle-|VVH\rangle)$, $|GHZ_3^-\rangle=\frac{1}{\sqrt{2}}(|HVH\rangle-|VHV\rangle)$ and $|GHZ_4^-\rangle=\frac{1}{\sqrt{2}}(|VHH\rangle-|HVV\rangle)$ can cause Bob to obtain incorrect raw keys (see the Appendix \ref{Appendix A}). Therefore, the QBER $Q_{1}$ caused by the decoherence in the white-noise model is given by
\begin{eqnarray}\label{Q}
Q_{1}=4\frac{1-F}{8}\eta^3=\frac{1-F}{2}\eta^3.
\end{eqnarray}

Next we analyze the impact of the photon loss on the DI QSS protocol. We assume that each user adopts a three-value strategy for the measurements. In detail, besides the measurement results $+1$ ($|+_x\rangle)$ and $-1$ $(|-_x\rangle)$, each user defines an extra output $\bot$ for the no-click event. For simplicity, we denote the measurement results $|+_x\rangle$, $|-_x\rangle$, and no-click event by $+$, $-$, and $\bot$, respectively. After the entanglement distribution, the probability of the measurement results $P(a_{i}b_{j}c_{k})$ has the following four cases (we do not consider the decoherence here): (i) no photon loss,
\begin{eqnarray}\label{0lost}
P(+++)&=&P(+--)=P(-+-)=P(--+) \nonumber\\
&=&\left(\frac{1}{2}\eta\right)\left(\frac{1}{2}\eta\right)\eta=\frac{1}{4}\eta^3,\nonumber\\
P(++-)&=&P(+-+)=P(-++)=P(---)=0;\nonumber\\
\end{eqnarray}
(ii) one photon is lost,
\begin{eqnarray}\label{1lost}
&&P(++\bot)=P(+-\bot)=P(+\bot+)=P(+\bot-)\nonumber\\
&&=P(-+\bot)=P(--\bot)=P(-\bot+)=P(-\bot-)\nonumber\\
&&=P(\bot++)=P(\bot+-)=P(\bot-+)=P(\bot--)\nonumber\\
&&=\left(\frac{1}{2}\eta\right)\left(\frac{1}{2}\eta\right)\bar{\eta}=\frac{1}{4}\eta^{2}\bar{\eta};
\end{eqnarray}
(iii) two photons are lost,
\begin{eqnarray}\label{2lost}
&&P(+\bot\bot)=P(-\bot\bot)=P(\bot+\bot)=P(\bot-\bot)\nonumber\\
&=&P(\bot\bot+)=P(\bot\bot-)=\left(\frac{1}{2}\eta\right)\bar{\eta}\bar{\eta}=\frac{1}{2}\eta\bar{\eta}^{2};
\end{eqnarray}
and (iv) all three photons are lost,
\begin{eqnarray}\label{3lost}
P(\bot\bot\bot)=\bar{\eta}^{3}.
\end{eqnarray}

If the users obtain cases (ii)-(iv), then by definition an error occurs. As a result, the QBER caused by the photon loss can be calculated as
\begin{eqnarray}
Q_{2}&=&12\left(\frac{1}{4}\eta^{2}\bar{\eta}\right)+6\left(\frac{1}{2}\eta\bar{\eta}^{2}\right)+\bar{\eta}^{3}\nonumber\\
&=&3\eta^2\bar{\eta}+3\bar{\eta}^2\eta+\bar{\eta}^3=1-\eta^{3}.
\end{eqnarray}
Combining the decoherence and photon loss, the total QBER $\delta$ can be calculated as
\begin{eqnarray}\label{lossdelta}
\delta&=&Q_{1}+Q_{2}=\frac{1-F}{2}\eta^3+1-\eta^{3}\nonumber\\
&=&1-\frac{1}{2}\eta^3-\frac{1}{2}\eta^3F.
\end{eqnarray}

Therefore, the key error rate $H(A_1|B_1,C_1)$  is given by
\begin{eqnarray}\label{HA1B1}
H\left(A_1|B_1,C_1\right)=h\left(\delta\right),
\end{eqnarray}
where $h(x)$ is the binary Shannon entropy with $h(x)=-x\log_{2}{x}-(1-x)\log_{2}{(1-x)}$.

Based on the form of $\rho_{ABC}$, the theoretical value of the CHSH polynomial between Alice's and Bob's measurement results is given by
\begin{eqnarray}\label{losssm}
S=2\sqrt{2}F\eta^3=2\sqrt{2}(2-\eta^{3}-2\delta).
\end{eqnarray}

For the collective attack model, we suppose that Eve intercepts the distributed GHZ state and couples the quantum state with her photon. The quantum state of the whole system can be written as $\rho_{ABCE}^{\otimes N}=|\Psi_{ABCE}\rangle^{\otimes N}\langle\Psi_{ABCE}|$, where $|\Psi_{ABCE}\rangle=\frac{1}{\sqrt{2}}(|HHH\rangle_{ABC}|\psi_{H}\rangle_{E}+|VVV\rangle_{ABC}|\psi_{V}\rangle_{E})$. Then Eve distributes the three subsystems to Alice, Bob, and Charlie, and retains $\rho_E=Tr_{ABC}[\rho_{ABCE}]$ for herself. In DI QSS, Charlie has to announce his subkeys to cooperate with Bob. In this way, Eve can obtain Charlie's subkeys. The above states saturate the lower bound of $H(A_1|E)$ as
\begin{eqnarray}\label{HA1E}
H(A_1|E)\geq 1-h\left(\frac{\sqrt{S^2/4-1}}{2}+\frac{1}{2}\right).
\end{eqnarray}

By substituting Eqs.~\eqref{lossdelta}-\eqref{HA1E} into Eq.~\eqref{keyrate}, we can obtain a lower bound of the key generation rate $r$ as a function of the global detection efficiency $\eta$ and the fidelity $F$ as
\begin{eqnarray}\label{lossrate}
r&\ge& 1-h\left(\frac{\sqrt{2F^2\eta^6-1}}{2}+\frac{1}{2}\right)\nonumber\\
&-&h\left(1-\frac{1}{2}\eta^3-\frac{1}{2}\eta^3F\right).
\end{eqnarray}

The key generation rate $r$ as a function of the global detection efficiency $\eta$ is shown in Fig. \ref{fig:boat0}. Here the fidelity of the target GHZ state is $F=1, 0.99, 0.97, 0.95$. Both the decoherence and photon loss seriously reduce DI QSS's key generation rate. To obtain a positive value of $r$, high global detection efficiency  and fidelity are required. When $F=1$ (no decoherence), the global detection efficiency threshold of the DI QSS protocol is $96.32\%$. When the fidelity $F$ decreases, DI QSS requires higher global detection efficiency. For example, when $F=0.95$, the global detection efficiency threshold increases to $97.57\%$. Meanwhile, the tolerable threshold of $F$ is 85.1\%. Below the global detection efficiency threshold or fidelity threshold, no positive value of $r$ can be achieved.

\begin{figure}
\includegraphics[scale=0.72]{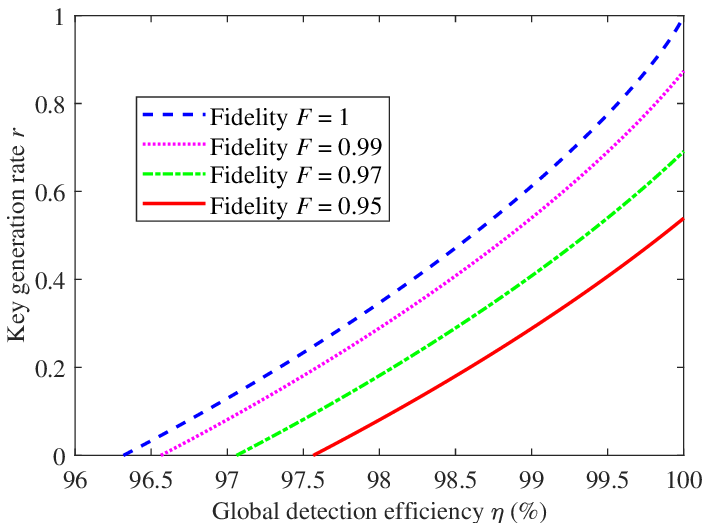}
\caption{(Color online) Key generation rate $r$ of our DI QSS protocol as a function of the global detection efficiency $\eta$. Here the fidelity of the target GHZ state is $F=1, 0.99, 0.97, 0.95$.}
\label{fig:boat0}
\end{figure}

\section{Active improvement strategies}\label{Section4}
From the calculations in Sec. \ref{Section3}, the DI QSS protocol has quite a high global detection efficiency requirement and low noise resistance, which would largely increase the experiment's difficulty. To address the above defects, we introduce the noise preprocessing strategy and postselection strategy in our DI QSS protocol. We refer to the active operation methods as active improvement strategies.

\subsection{DI QSS protocol with noise preprocessing strategy}\label{Section4.1}
The noise preprocessing strategy, which is implemented by adding some artificial noise to the initial measurement data, has been adopted in DI QKD protocols to enhance the noise resistance \cite{DIQKD8,DIQKD9,DIQKD10}. Here, Alice performs the preprocessing operation in Step 3 of the DI QSS protocol. In detail, when the measurement basis combination is $\{A_{1}B_{1}C_{1}\}$, Alice flips her measurement result with a probability of $q$ (flips $+1$ to $-1$, and $-1$ to $+1$). The additional noise damages both the correlation between Alice's and Bob's key bits and the correlation between Alice's and Eve's key bits. Since the possibility to generate a key depends on the difference between these two correlations, the net effect can  be positive. After the measurements and flip operations for all photons, Alice announces the flip probability $q$ in the error correction process. Then the users apply a hash function to the raw keys to obtain the final secret keys. The security of our DI QSS protocol with noise preprocessing strategy is similar as that of the DI QKD with noise preprocessing \cite{DIQKD8,DIQKD9}.

By performing the noise preprocessing strategy, the total noise quantum bit error rate $\delta_q$ consists of two scenarios. First, the initial measurement results do not suffer from the bit-flip error but Alice flips her result with a probability of $q$. Second, the initial results suffer from the bit-flip error and Alice does not flip her result with a probability of $1-q$. Therefore, $\delta_q$ is given by
\begin{eqnarray}\label{noisepredelta}
\delta_q=q\left(1-\delta\right)+\left(1-q\right)\delta=q+\left(1-2q\right)\delta.
\end{eqnarray}
In this way, the key error rate will change to
\begin{eqnarray}\label{HqA1B1}
H\left(A_1|B_1,C_1\right)_q=h\left(\delta_q\right).
\end{eqnarray}
 It is naturally that $H\left(A_1|B_1,C_1\right)_q$ will increase with the growth of $q$.

Based on the derivation in the DI QKD protocols with the noise preprocessing strategy \cite{DIQKD8,DIQKD9,DIQKD10},
the key secrecy rate to Eve after the noise preprocessing operation can be lower bounded by
\begin{eqnarray}\label{HqA1E}
H\left(A_1|E\right)_q\geq g\left(S,q \right),
\end{eqnarray}
where
\begin{eqnarray}\label{gsm}
g\left(S,q\right)&=&1-h\left(\frac{\sqrt{\frac{S^2}{4}-1}}{2}+\frac{1}{2}\right)\nonumber\\
&+&h\left[\frac{\sqrt{\left(1-2q\right)^2+4q\left(1-q\right)\left(\frac{S^2}{4}-1\right)}}{2}+\frac{1}{2}\right].\nonumber\\
\end{eqnarray}
The last item of $g\left(S,q \right)$ makes $H\left(A_1|E\right)_q$ higher than $H\left(A_1|E\right)$ in Eq.~\eqref{HA1E}. As a result, the noise preprocessing operation can reduce the key leakage rate. Finally, by substituting  Eqs.~\eqref{HqA1B1}-\eqref{gsm} into Eq.~\eqref{keyrate}, we can obtain the lower bound of the key generation rate  of the DI QSS protocol with the noise preprocessing strategy as
\begin{eqnarray}\label{rateq}
r_q\ge g\left[2\sqrt{2}\left(2-\eta^3-2\delta\right),q\right]-h\left[q+\left(1-2q\right)\delta\right].
\end{eqnarray}

\begin{figure}
\includegraphics[scale=0.77]{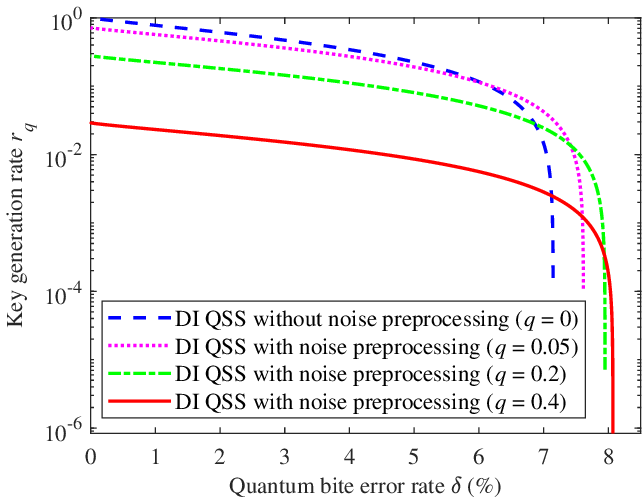}
\caption{(Color online) Key generation rate $r_q$ of the DI QSS protocol with the noise preprocessing strategy as a function of the QBER $\delta$ for different values of $q$.}
\label{fig:boat2}
\end{figure}

In Fig. \ref{fig:boat2} we provide a plot of $r_{q}$ versus the QBER $\delta$ in Eq.~\eqref{lossdelta} for different values of $q$. Here we fix the global detection efficiency $\eta=1$ and adjust the flipping probability $q=0,0.05,0.2,0.4$. It can be found that the noise tolerable threshold of $\delta$ for the DI QSS without the noise preprocessing strategy ($q=0$) is 7.148\%. When Alice sets $q=0.05, 0.2, 0.4$, the noise tolerable threshold can be increased to 7.616\%, 7.95\%, and 8.072\%, respectively. As a result, the adoption of a noise preprocessing strategy can enhance the noise resistance of the DI QSS protocol. However, the adoption of a noise preprocessing strategy would reduce $r_{q}$ since it increases the key error rate to Bob. For example, when Alice sets $q=0.4$ and $\delta=0.5$, the key generation rate is only about 3.8\% of that in the DI QSS protocol without the noise preprocessing strategy.

\subsection{DI QSS protocol with postselection strategy}
In this section, three users all adopt the postselection strategy in the measurement process to enhance DI QSS's photon loss resistance. The postselection method takes all the click events into account, including the no-click events.  The security of the DI QKD with the postselection strategy has been proved \cite{DIQKD9,DIQKD12}. The security of our DI QSS protocol with the postselection strategy can be also proved in a similar way. In step 2, each user replaces the previous three-value strategy with a deterministic two-value strategy. In detail, besides the deterministic results $+1$ and $-1$, each user labels a no-click result as $+1$. Based on Eqs.~\eqref{0lost}-\eqref{3lost}, the postselection strategy maps the detector response cases to
\begin{eqnarray}\label{Bmapping}
\left\{++\bot\right\},\left\{+\bot+\right\},\left\{\bot++\right\},\nonumber\\
\left\{+\bot\bot\right\},\left\{\bot+\bot\right\},\left\{\bot\bot+\right\},\left\{\bot\bot\bot\right\}&\mapsto&\left\{+++\right\},\nonumber\\
\left\{\bot--\right\}&\mapsto&\left\{+--\right\},\nonumber\\
\left\{-\bot-\right\}&\mapsto&\left\{-+-\right\},\nonumber\\
\left\{--\bot\right\}&\mapsto&\left\{--+\right\},\nonumber\\
\left\{+\bot-\right\},\left\{\bot+-\right\},\left\{\bot\bot-\right\}&\mapsto&\left\{++-\right\},\nonumber\\
\left\{+-\bot\right\},\left\{\bot-+\right\},\left\{\bot-\bot\right\}&\mapsto&\left\{+-+\right\},\nonumber\\
\left\{-+\bot\right\},\left\{-\bot+\right\},\left\{-\bot\bot\right\}&\mapsto&\left\{-++\right\}.\nonumber\\
\end{eqnarray}

With the postselection strategy, the probability of the measurement result changes to
\begin{eqnarray}\label{BP1}
P_p\left(+++\right)&=&P\left(+++\right)+P\left(++\bot\right)+P\left(+\bot+\right)\nonumber\\
&+&P\left(\bot++\right)+P\left(+\bot\bot\right)+P\left(\bot+\bot\right)\nonumber\\
&+&P\left(\bot\bot+\right)+P\left(\bot\bot\bot\right),\nonumber\\
P_p\left(+--\right)&=&P\left(+--\right)+P\left(\bot--\right),\nonumber\\
P_p\left(-+-\right)&=&P\left(-+-\right)+P\left(-\bot-\right),\nonumber\\
P_p\left(--+\right)&=&P\left(--+\right)+P\left(--\bot\right),\nonumber\\
P_p\left(++-\right)&=&P\left(++-\right)+P\left(+\bot-\right)+P\left(\bot+-\right)\nonumber\\
&+&P\left(\bot\bot-\right),\nonumber\\
P_p\left(+-+\right)&=&P\left(+-+\right)+P\left(+-\bot\right)+P\left(\bot-+\right)\nonumber\\
&+&P\left(\bot-\bot\right),\nonumber
\end{eqnarray}
\begin{eqnarray}
P_p\left(-++\right)&=&P\left(-++\right)+P\left(-+\bot\right)+P\left(-\bot+\right)\nonumber\\
&+&P\left(-\bot\bot\right),\nonumber\\
P_p\left(---\right)&=&P\left(---\right).
\end{eqnarray}

Therefore, from Eqs.~\eqref{Q}-\eqref{3lost} and \eqref{BP1}, the total QBER is given by
\begin{eqnarray}\label{postdelta}
\delta_p&=&Q_{1}+P_p(++-)+P_p(+-+)+P_p(-++)\nonumber\\
&+&P_p(---)\nonumber\\
&=&\frac{1-F}{2}\eta^3+\left(\frac{1}{2}\eta\right)\left(\frac{1}{2}\eta\right)6\bar{\eta}+\left(\frac{1}{2}\eta\right)3\bar{\eta}\bar{\eta}\nonumber\\
&=&\frac{1-F}{2}\eta^3-\frac{3}{2}\eta^2+\frac{3}{2}\eta.
\end{eqnarray}
Compared with $\delta$ in Eq.~\eqref{lossdelta}, the adoption of postselection strategy can reduce the error rate caused by the photon loss.

We can obtain the key error rate as
\begin{eqnarray}\label{HpA1B1}
H\left(A_1|B_1,C_1\right)_p=h\left(\delta_p\right).
\end{eqnarray}

On the other hand, the adoption of the postselection would change the CHSH polynomial. The theoretical value of the CHSH polynomial between Alice and Bob changes to
\begin{eqnarray}\label{postsm}
S_{p}=2\sqrt{2}\eta^3F+2\bar{\eta}^3,
\end{eqnarray}
which is higher than the original value of $S$ in Eq. (\ref{losssm}).

Substituting Eq.~\eqref{postsm} into Eq.~\eqref{HA1E}, we can  obtain the key secrecy rate to Eve as
\begin{eqnarray}\label{HpA1E}
H\left(A_1|E\right)_p&\geq&1-h\left(\frac{\sqrt{S_{p}^2/4-1}}{2}+\frac{1}{2}\right)\nonumber\\
&=&1-h\left[\frac{\sqrt{\left(\sqrt{2}\eta^3F+\bar{\eta}^3\right)^2-1}}{2}+\frac{1}{2}\right].\nonumber\\
\end{eqnarray}
It can be found that with the higher value of the CHSH polynomial, the key secrecy rate to Eve can be effectively increased.

Finally, a lower bound of the DI QSS's key generation rate $r_p$ with the postselection strategy can be derived as
\begin{eqnarray}\label{ratep}
r_p\ge1&-&h\left[\frac{\sqrt{\left(\sqrt{2}\eta^3F+\bar{\eta}^3\right)^2-1}}{2}+\frac{1}{2}\right]\nonumber\\
&-&h\left(\frac{1-F}{2}\eta^3-\frac{3}{2}\eta^2+\frac{3}{2}\eta\right).
\end{eqnarray}

\begin{figure}
\includegraphics[scale=0.75]{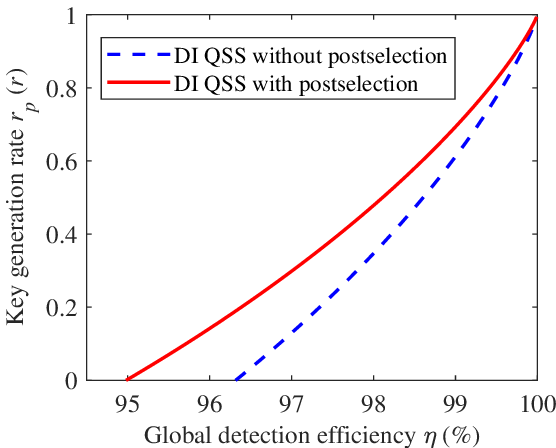}
\caption{(Color online) Key generation rate $r_p$ ($r$) of the DI QSS protocol with (without) the postselection strategy as a function of the global detection efficiency $\eta$. Here the fidelity $F=1$.}
\label{fig:boat3}
\end{figure}

In Fig. \ref{fig:boat3} we analyze the key generation rate $r_p$ ($r$) of our DI QSS protocol with (without) the postselection strategy versus the global detection efficiency $\eta$. Here we set the fidelity $F=1$ for simplicity. By performing the postselection strategy, the global detection efficiency threshold can be reduced from 96.32\% to 94.99\%. This improvement enhances DI QSS's photon loss resistance. Meanwhile, benefitting from the postselection strategy, the key generation rate of DI QSS can also be effectively increased.

\subsection{The DI QSS protocol with advanced postselection strategy}
In this section we propose an advanced postselection strategy which combines the noise preprocessing and the postselection to further improve the performance of the DI QSS protocol.
Based on Eqs.~\eqref{noisepredelta} and \eqref{postdelta}, with the advanced postselection strategy, the total QBER  of the DI QSS protocol can be derived as
\begin{eqnarray}\label{deltaqp}
\delta_{qp}&=&q+\left(1-2q\right)\delta_p\nonumber\\
&=&q+(1-2q)(\frac{1-F}{2}\eta^3-\frac{3}{2}\eta^2+\frac{3}{2}\eta).
\end{eqnarray}

Therefore, the key error rate has the form
\begin{eqnarray}\label{HqpA1B1}
H\left(A_1|B_1,C_1\right)_{qp}=h\left(\delta_{qp}\right).
\end{eqnarray}

As the noise preprocessing would not influence the CHSH polynomial, the CHSH polynomial $S_{qp}$ has the same form as $S_{p}$ in Eq.~\eqref{postsm}. By substituting $S$ in Eq.~\eqref{HqA1E} with $S_{p}$, the key secrecy rate to Eve is lower bounded by
\begin{eqnarray}
H\left(A_1|E\right)_{qp}\geq g\left(2\sqrt{2}\eta^3F+2\bar{\eta}^3,q\right).
\end{eqnarray}

Therefore, we can derive the lower bound of the key generation rate  with the advanced postselection strategy as
\begin{eqnarray}\label{rateqp}
r_{qp}&\ge& g\left(2\sqrt{2}\eta^3F+2\bar{\eta}^3,q\right)\nonumber\\
&-&h\left[q+(1-2q)(\frac{1-F}{2}\eta^3-\frac{3}{2}\eta^2+\frac{3}{2}\eta)\right].
\end{eqnarray}

\begin{figure}
\includegraphics[scale=0.75]{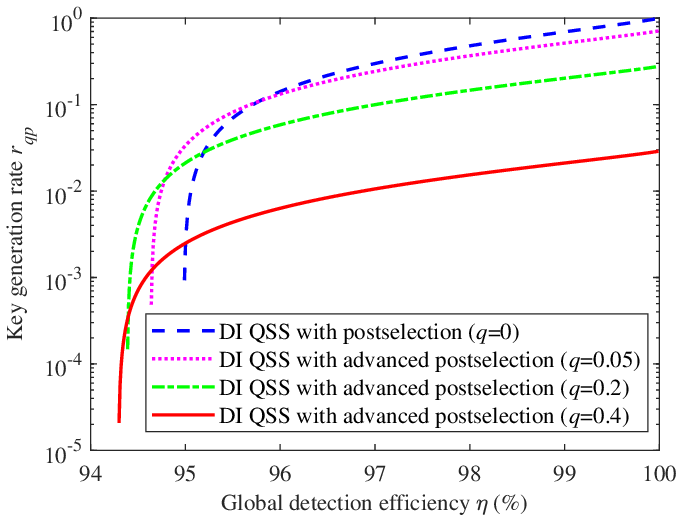}
\caption{(Color online) Key generation rate $r_{qp}$ as a function of the global detection efficiency $\eta$  with $q=0,0.05,0.2,0.4$. Here the fidelity $F=1$.}
\label{fig:boat4}
\end{figure}

Figure \ref{fig:boat4} shows a plot of the key generation rate $r_{qp}$, with $q=0, 0.05,0.2,0.4$ versus the global detection efficiency $\eta$, with the fidelity $F=1$. It can be found that the advanced postselection strategy can further relax the global detection efficiency threshold. When $q$ increases from 0 to 0.4, the global detection efficiency threshold is further reduced from 94.99\% to 94.3\%. It has been proved that the DI QSS with the advanced postselection strategy has higher photon loss resistance. However, similar to the results in Sec. \ref{Section4.1}, the higher photon loss resistance will sacrifice the key generation rate. With the growth of $q$, the key generation rate will be reduced.

The photon transmission efficiency in the noisy quantum channel can be calculated as $\eta_t=10^{\alpha d/10}$, where $d$ represents the photon transmission distance and $\alpha=0.2$ dB/km for a standard optical fiber. We define the detection efficiency of each photon detector as $\eta_{d}$ and the coupling efficiency of the photon to the fiber as $\eta_{c}$. In this way, the global detection efficiency $\eta$ can be obtained as $\eta=\eta_{t}\eta_{d}\eta_{c}$. It has been reported that the superconducting nanowire single-photon detectors with ultralow dark counts can achieve detection efficiencies over 90\%, even over 98\% at 1550 nm \cite{detector0,SNSPD,detector1}. In this way, we suppose  $\eta_d=98\%$ and $\eta_{c}=99\%$ for the simulation.

Figure \ref{fig:boat7} shows the key generation rate of the DI QSS protocol in four scenarios as a function of the photon transmission distance $d$, i.e., the DI QSS protocol without any active improvement strategy, with the noise preprocessing strategy ($q=0.2$), with the postselection strategy, and with the advanced postselection strategy ($q=0.2$). It can be seen that the DI QSS protocol with the advanced postselection strategy ($q=0.2$) has the longest photon transmission distance threshold of 0.59 km. In this way, the maximal secure communication distance between any two users is about 1.02 km. In addition, the DI QSS protocol with the postselection strategy has the highest key generation rate. For example, with $d=0.3$ km, the key generation rate in this scenario is about 2 times that in the DI QSS protocol with the advanced postselection strategy ($q=0.2$). As a result, in future applications, we should consider both the key generation rate and the communication distance factors, and select the optimal active improvement strategy.

\begin{figure}
\includegraphics[scale=0.76]{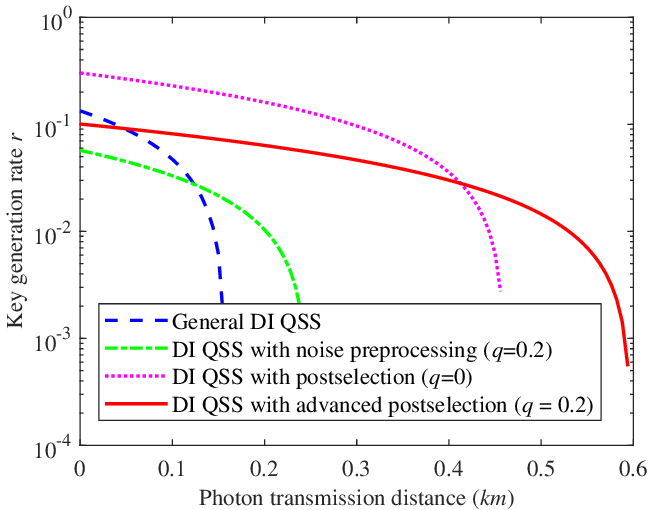}
\caption{(Color online) Key generation rate of the DI QSS protocol in four scenarios as a function of the photon transmission distance $d$. Here, we control the detection efficiency  $\eta_d=98\%$, the coupling efficiency $\eta_{c}=99\%$, and the fidelity $F=1$.}
\label{fig:boat7}
\end{figure}

\section{Discussion and conclusion}\label{Section5}
We have proposed a DI QSS protocol with noise preprocessing and postselection and estimated its performance in practical communication scenarios. Our DI QSS protocol requires the distribution of the three-photon GHZ state to three parties through quantum channels. The security of our DI QSS protocol is based on the measurement results violating the  Svetlichny inequality. The generation of three-photon GHZ states with high fidelity has been extensively studied \cite{GHZ0,GHZ1,GHZ2,GHZ3,GHZ4}. Hamel \emph{et al.} \cite{GHZ2} eliminated the limitation of the outcome postselection by cascading two SPDC  sources for the direct generation of three-photon polarization GHZ states (as shown in Fig. \ref{fig:boat1}). The fidelity $F_s$ of the target GHZ states reached 86\%. Later, with the phase-stable source, the fidelity of the target GHZ state increased to over 96\% \cite{GHZ4}. In the practical GHZ state generation, only the phase-flip error may occur \cite{GHZ2,GHZ4}, which will cause the quantum bit error in the key generation process. In this way, the GHZ state generation in Ref. \cite{GHZ4} would cause a QBER of 4\%. This QBER is below the noise threshold of the DI QSS protocol, so the current GHZ state generation technology in Ref. \cite{GHZ4} can meet the requirement of our DI QSS protocol.

\begin{figure}
\includegraphics[scale=0.8]{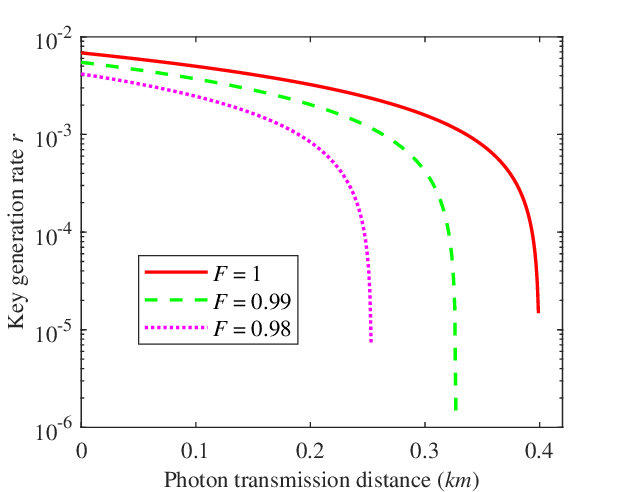}
\caption{(Color online) Key generation rate of the DI QSS protocol with the advanced postselection strategy as a function of the photon transmission distance $d$. Here, we control the detection efficiency  $\eta_d=98\%$ , the coupling efficiency $\eta_{c}=99\%$, and the fidelity $F=1,0.99,0.98$.}
\label{fig:boat8}
\end{figure}

In Secs.~\ref{Section3} and \ref{Section4} we considered the perfect GHZ state source. We noted that the practical imperfect GHZ state source will naturally decrease the noise tolerance threshold of the entanglement distribution process and the secure communication distance. We assumed that the GHZ state source has a fidelity of $F_s=96\%$. The total QBER caused by the imperfect GHZ state source and the decoherence could be calculated as $\delta=F_s \frac{1-F}{2}+(1-F_{s})F+(1-F_s)\frac{1-F}{2}=\frac{1}{2}+\frac{1}{2}F-F_{s}F$. Figure \ref{fig:boat8} shows the key generation rate of the DI QSS protocol with the advanced postselection strategy as a function of the photon transmission distance $d$ for the imperfect GHZ state source. We set the fidelity $F=1,0.99,0.98$. Since the total QBER threshold was $8.072\%$, we could obtain a threshold of $F$ of 91.14\%, corresponding to the noise tolerance threshold during the photon transmission process of 4.426\% and the maximal secure communication distance between any two users of 0.693 km.

The high global detection efficiency requirement and low noise resistance are two main obstacles for DI QSS's experimental demonstration. Compared with the QKD protocol with a global detection efficiency of 92.4\% \cite{DIQKD1,DIQKD2}, our DI QSS protocol requires a higher global detection efficiency of 96.32\%. The reason is that DI QSS has to distribute multi-photon entanglement to multiple parties through noisy quantum channels. The photon loss occurring in any photon transmission process would destroy the entanglement. As a result, DI QSS is more vulnerable to photon loss and the maximal secure distance between any two users is short (about 0.26 km). With the active improvement strategies, we can effectively relax the global detection efficiency threshold and increase the noise threshold of DI QSS. In detail, with the same level of noise preprocessing strategy ($q=0.4$) adopted, the noise tolerance threshold of DI QSS can be improved to 8.072\% (a relative improvement of 12.92\%) with the perfect GHZ state source, which is comparable to that of DI QKD. Moreover, by combining the noise preprocessing and postselection strategies, the global detection efficiency threshold of our DI QSS protocol with the perfect GHZ state source is reduced to 94.29\% ($q\rightarrow0.5$) and the maximal secure distance between any two users is increased to about 1.06 km, which is about 4 times of the original value. The above improvement can effectively facilitate the experimental demonstration of DI QSS. In \cite{SNSPD}, a superconducting nanowire single-photon detector with 98\% system detection efficiency at 1550 nm was experimentally realized, technology which enables realization of our DI QSS protocol with high global detection efficiency. The value of $\eta_c$ also influences the global detection efficiency. As $\eta_c$ decreases, the key generation rate and secure communication distance will be reduced. The threshold of $\eta_c$ is calculated as $97\%$.

Based on the research of DI QKD \cite{DIQKD9,DIQKD10}, there are some other possible approaches to relaxing the requirements for experimental devices and improving DI QSS's performance such as the adoption of the partially entangled GHZ states and optimal measurements. Meanwhile, performing the quantum heralded amplification \cite{NLA0,NLA1,NLA2} and GHZ state entanglement purification \cite{EPP0,EPP1,EPP2} after the photon transmission may also enhance the three-photon entanglement. Moreover, the quantum repeater is a promising method for building a long-distance entanglement channel and quantum network \cite{repeater1,repeater2}. Combining DI QSS with the quantum repeater may be a promising way to further increase the secure communication distance. Note that our work considered the case where the device is used with $N$ ($N\rightarrow\infty$) rounds and estimates the asymptotic key generation rate. In practical experiments, we addressed the question of what sample size is needed for the DI QSS security proof, which is called the finite-size key problem. Actually, the finite-size security proof of the DI-type protocol is an independent security proof, which replies on the entropy accumulation theorem and large-scale semidefinite programming to estimate Eve's uncertainty of keys \cite{DIQKD15}. The combination of the finite-size effect, the noise preprocessing and the postselection strategies in DI QSS is an interesting issue left for future work.

In this paper we considered the DI QSS with three users based on the violation of the three-party Svetlichny inequality. The $m$-party Svetlichny inequality was proposed in \cite{nonlocality7} and can be written as
\begin{eqnarray}\label{mSvetlichny}
S_{m}=\langle S_{m-1}M_{2}\rangle+\langle S_{m-1}'M_{1}\rangle \leq 2^{m-1}.
\end{eqnarray}
When $m=3$, Eq.~\eqref{mSvetlichny} degenerates into the tripartite Svetlichny inequality \eqref{Svetlichnypolynomial2}. When $m=4$, Eq.~\eqref{mSvetlichny} is equivalent to $S_{ABCD}=\langle S_{ABC}D_{2}\rangle+\langle S_{ABC}'D_{1}\rangle\leq 8$. The violation of the $m$-party Svetlichny inequality indicates the nonlocal correlation among the particles in $m$ users. Making use of the $m$-party Svetlichny inequality violation, our DI QSS can be generalized to the arbitrary $m$-user situation.

In conclusion, QSS is a fundamental quantum secure communication primitive. In theory, DI QSS can resist all possible attacks focusing on imperfect experimental devices and thus provide the highest security for QSS under practical imperfect experimental conditions. The original DI QSS protocol proved its correctness and completeness under a causal independence assumption regarding measurement devices. However, there has been a lack of DI QSS's performance characterization in practical communication situation, which largely impedes its experimental demonstration and application in the future quantum secure communication field. Here we proposed a DI QSS protocol with noise preprocessing and postselection (active improvement strategies). The security of the DI QSS protocol is based on users' measurement results violating the Svetlichny inequality. We researched its performances in practical communication situations without and with the active improvement strategies, including the key generation rate via the von Neumann entropy, the global detection efficiency, the noise threshold, and the maximal communication distance between any two users. Our DI QSS has two advantages. First, it is a DI QSS protocol in a practical communication situation, where we developed numerical methods to estimate its practical performance. Second, we adopted the active improvement strategy in the DI QSS protocol, which can reduce DI QSS's global detection efficiency threshold from 96.32\% to 94.30\% and increase the noise threshold  from 7.148\% to 8.072\%. These improvements can  promote DI QSS's experimental demonstration and applications in the future.

\appendix
\section{The white noise model for GHZ States}\label{Appendix A}
\setcounter{equation}{0}
\setcounter{subsection}{0}
\renewcommand{\theequation}{A.\arabic{equation}}

In the white-noise model, Alice, Bob, and Charlie share a noisy GHZ state in the form
\begin{eqnarray}\label{BGHZ}
\rho_{ABC}=F|GHZ\rangle\langle GHZ|+\frac{1-F}{8} I,
\end{eqnarray}
where the fidelity $F$ is the probability that the photon state is free of errors. The unit matrix $I$ consists of a density matrix of eight GHZ states as
\begin{eqnarray}\label{BGHZ2}
I&=&|GHZ_1^+\rangle\langle GHZ_1^+|+|GHZ_1^-\rangle\langle GHZ_1^-| \nonumber\\
&+&|GHZ_2^+\rangle\langle GHZ_2^+|+|GHZ_2^-\rangle\langle GHZ_2^-|\nonumber\\
&+&|GHZ_3^+\rangle\langle GHZ_3^+|+|GHZ_3^-\rangle\langle GHZ_3^-|\nonumber\\
&+&|GHZ_4^+\rangle\langle GHZ_4^+|+|GHZ_4^-\rangle\langle GHZ_4^-|,
\end{eqnarray}
where,
\begin{eqnarray}\label{BGHZ3}
|GHZ_1^{\pm}\rangle&=&\frac{1}{\sqrt{2}}\left(|HHH\rangle \pm |VVV\rangle\right),\nonumber\\
|GHZ_2^{\pm}\rangle&=&\frac{1}{\sqrt{2}}\left(|HHV\rangle \pm |VVH\rangle\right),\nonumber\\
|GHZ_3^{\pm}\rangle&=&\frac{1}{\sqrt{2}}\left(|HVH\rangle \pm |VHV\rangle\right),\nonumber\\
|GHZ_4^{\pm}\rangle&=&\frac{1}{\sqrt{2}}\left(|VHH\rangle \pm |HVV\rangle\right).
\end{eqnarray}

In the white-noise model, the target GHZ state ($|GHZ_1^+\rangle$) will degrade to one of the states in Eq.~\eqref{BGHZ3} with equal probability $\frac{1-F}{8}$. After the measurement with the basis combination of  $\{A_{1}B_{1}C_{1}\}$, the measurement results of the eight GHZ states can be written as
\begin{widetext}
\begin{eqnarray}\label{BGHZ4}
|GHZ_1^+\rangle&=&\frac{1}{2}(|+_x\rangle|+_x\rangle|+_x\rangle+|+_x\rangle|-_x\rangle|-_x\rangle+|-_x\rangle|+_x\rangle|-_x\rangle+|-_x\rangle|-_x\rangle|+_x\rangle),\nonumber\\
|GHZ_2^+\rangle&=&\frac{1}{2}(|+_x\rangle|+_x\rangle|+_x\rangle-|+_x\rangle|-_x\rangle|-_x\rangle-|-_x\rangle|+_x\rangle|-_x\rangle+|-_x\rangle|-_x\rangle|+_x\rangle),\nonumber\\
|GHZ_3^+\rangle&=&\frac{1}{2}(|+_x\rangle|+_x\rangle|+_x\rangle-|+_x\rangle|-_x\rangle|-_x\rangle+|-_x\rangle|+_x\rangle|-_x\rangle-|-_x\rangle|-_x\rangle|+_x\rangle),\nonumber\\
|GHZ_4^+\rangle&=&\frac{1}{2}(|+_x\rangle|+_x\rangle|+_x\rangle+|+_x\rangle|-_x\rangle|-_x\rangle-|-_x\rangle|+_x\rangle|-_x\rangle-|-_x\rangle|-_x\rangle|+_x\rangle),\nonumber\\
|GHZ_1^-\rangle&=&\frac{1}{2}(|+_x\rangle|-_x\rangle|+_x\rangle+|-_x\rangle|+_x\rangle|+_x\rangle+|+_x\rangle|+_x\rangle|-_x\rangle+|-_x\rangle|-_x\rangle|-_x\rangle),\nonumber\\
|GHZ_2^-\rangle&=&\frac{1}{2}(|+_x\rangle|-_x\rangle|+_x\rangle+|-_x\rangle|+_x\rangle|+_x\rangle-|+_x\rangle|+_x\rangle|-_x\rangle-|-_x\rangle|-_x\rangle|-_x\rangle),\nonumber\\
|GHZ_3^-\rangle&=&\frac{1}{2}(-|+_x\rangle|-_x\rangle|+_x\rangle+|-_x\rangle|+_x\rangle|+_x\rangle+|+_x\rangle|+_x\rangle|-_x\rangle-|-_x\rangle|-_x\rangle|-_x\rangle),\nonumber\\
|GHZ_4^-\rangle&=&\frac{1}{2}(|+_x\rangle|-_x\rangle|+_x\rangle-|-_x\rangle|+_x\rangle|+_x\rangle+|+_x\rangle|+_x\rangle|-_x\rangle-|-_x\rangle|-_x\rangle|-_x\rangle).
\end{eqnarray}
\end{widetext}

According to the coding rules in Sec. \ref{Section3.1}, the bit-flip error occurs when the photon state becomes $|GHZ_1^-\rangle$, $|GHZ_2^-\rangle$, $|GHZ_3^-\rangle$ or $|GHZ_4^-\rangle$.

\section*{Acknowledgement}
This work was supported by the National Natural Science Foundation of China under Grant Nos. 12175106 and 92365110, and the Natural Science Foundation of Jiangsu Province of China under Grant Nos. SBK2024047810 and SBK2024042659, and the Key R\&D Program of Guangdong Province under Grant No. 2018B030325002.

\end{document}